\documentclass[10pt]{iopart}

\usepackage{graphicx}
\usepackage{latexsym}
\usepackage{bbold}
\usepackage{bm}
\usepackage{color}
\usepackage{soul}
\usepackage{iopams}
\usepackage{diagbox}

\begin{document}

\title[Gyrokinetic prediction of core tungsten peaking]{Gyrokinetic prediction of core tungsten peaking in a WEST plasma with nitrogen impurities}

\author{J. Dominski}
\address{Princeton Plasma Physics Laboratory, 100 Stellarator rd, Princeton 08540 NJ, USA}
\ead{jdominsk@pppl.gov}
\author{P. Maget, P. Manas, J. Morales}
\address{CEA, IRFM, F-13108 Saint Paul-lez-Durance, France}
\author{S. Ku}
\address{Princeton Plasma Physics Laboratory, 100 Stellarator rd, Princeton 08540 NJ, USA}
\author{A. Scheinberg}
\address{Jubilee Development, Cambridge, MA, USA}
\author{C.S. Chang, R. Hager}
\address{Princeton Plasma Physics Laboratory, 100 Stellarator rd, Princeton 08540 NJ, USA}
\author{M. O'Mullane}
\address{UKAEA, Abingdon, United Kingdom of Great Britain and Northern Ireland}
\author{the WEST team}
\address{See http://west.cea.fr/WESTteam}

\vspace{10pt}
\begin{indented}
\item[]June 2024
\end{indented}

\begin{abstract}
Tungsten peaking is predicted in the core of a WEST plasma with total-f gyrokinetic simulations, including both collisional and turbulent transport. This prediction is validated with a synthetic diagnostic of the bolometry. Although nitrogen impurities are shown to reduce the neoclassical peaking of tungsten on-axis, the overall tungsten peaking increases when nitrogen impurities are present, as they reduce the turbulence screening off-axis. This finding is important for the plasma current ramp-up phase of ITER, where light impurities seeding will be desirable to achieve low temperatures at the plasma-facing components and reduce tungsten sputtering. It provides further argument for applying early ECRH heating to maintain margins on the core power balance. The neoclassical peaking factor is cross-verified between XGC and FACIT. The heat flux at separatrix and the heat load width are modeled by XGC and compared to WEST data.
\end{abstract}

\ioptwocol
\section{Introduction}
The WEST tokamak provides a unique platform for modeling integrated plasma scenarios within a full tungsten environment~\cite{Dittmar_2020,Bucalossi_2022}. 
Featuring an ITER-grade tungsten divertor, WEST is subject to tungsten contamination and associated radiative losses, which represent $\sim$50\% of the input power~\cite{goniche_NF2022}. This makes WEST particularly well-suited for studying the mechanisms leading to this resilient radiative fraction, providing valuable insights relevant to future ITER operations~\cite{DiGenova_NF2021}. 

Radiations can be very problematic during the ramp-up phase crossing the electron temperature ($\sim$1.5keV) that maximizes tungsten radiations. Injecting low-Z nitrogen impurities during this ramp-up phase was found to greatly improve the core temperature on WEST~\cite{Maget_2022}.  Indeed, radiations of low-Z impurities in the edge are known to reduce the tungsten contamination by reducing its sputtering from the wall~\cite{Kallenbach_2013}, but
profile effects and associated transport can also impact the radiative fraction by concentrating or removing tungsten impurities from the plasma core. 

With a dominant electron heating and no external torque WEST is particularly relevant for studying these transport mechanisms in a context relevant for ITER. 
For instance, during the radiation collapse, which occurs when the core electron heating is lacking, a reversal of the neoclassical pinch was shown to cause an accumulation of tungsten~\cite{Ostuni2022,Maget_PPCF2023} in plasmas without rotation. The tungsten peaking during these collapses was significantly lower than expected for a pure plasma, as explained by the presence of light impurities. This result motivated the present work on the influence of nitrogen impurities on the tungsten peaking. 

A total-f gyrokinetic study of the collisional and turbulent transport of tungsten impurities under the influence of nitrogen ions has thus been conducted with the gyrokinetic total-f code XGC~\cite{Ku16} in its version including multiple gyrokinetic ion species~\cite{Dominski19a,Dominski2024}. In addition, the turbulent heat load width on target is compared to experimental trends~\cite{Gaspar21} and 
the collisional peaking of tungsten is cross-verified between XGC and FACIT~\cite{Maget_PPCF2020b,Maget_PPCF2022,Fajardo_PPCF2022}.

The main result of the present work is the gyrokinetic prediction of the tungsten density peaking on axis which is validated against a synthetic diagnostic of the bolometer measurements~\cite{Devynck2021}. Nitrogen impurities are found to have a significant influence over the tungsten: they reduce the collisional peaking on axis but they also reduce the turbulent screening off-axis. Nitrogen stabilizing effect on turbulence is known~\cite{Bonanomi2018,Angioni2021,Maget_2022}, but here we reveal the influence of nitrogen on the competition between collisional and turbulent transport in determining the profile of tungsten peaking.

The reminder is organized as follows. 
Sec.~\ref{sec:west} describes the WEST plasma discharge \#55799 that is used for this  
study. 
Sec.~\ref{sec:XGC} introduces the electrostatic total-f model of XGC. 
{Sec.~\ref{sec:validation} presents a comparison between the transport modeled in the edge with XGC and WEST data.} 
Sec.~\ref{sec:col} presents a study of the collisional peaking factor of tungsten in presence of nitrogen impurities. 
Sec.~\ref{sec:turbulence} presents the follow-up study including turbulence. The core tungsten peaking is validated against a synthetic diagnostic of the bolometry. 
Finally a conclusion is drawn in Sec.~\ref{sec:conclusion}.
 
\section{XGC modeling of WEST plasma discharge \#55799}
\label{sec:west}
\begin{figure}
    \centering
    \includegraphics[width=8cm]{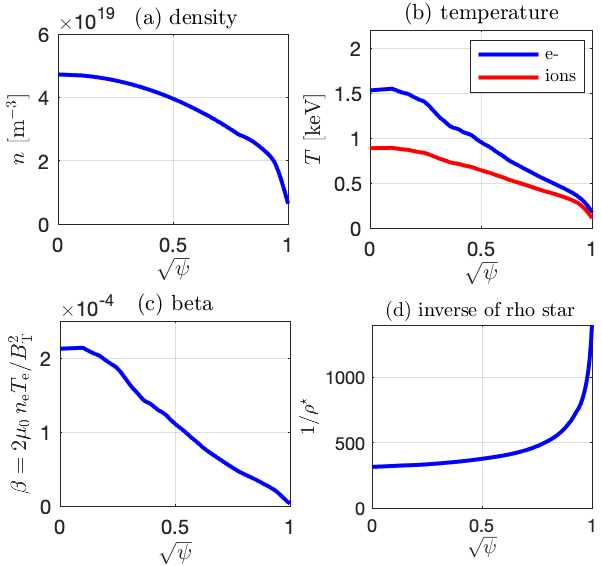}
    \caption{Profile of (a) density, (b) temperature, (c) beta parameter, and (d) aspect ratio $A=a/\rho_{\rm \rm th}$, for WEST shot number \#55799 at t=3.5s during the Ohmic phase. {In this manuscript, the radial direction is plotted with respect to the square-root of the normalized poloidal magnetic flux, $\sqrt{\psi}$}}
    \label{fig:profiles_nT_3.5s}
\end{figure}
The WEST plasma studied in this work is taken during the ohmic phase of discharge \#55799 at $t=3.5$s. {This L-mode plasma discharge was obtained at a standard operating magnetic field of B=3.7T, and plasma current 500 kA. At the time considered, the plasma shape has an elongation of 1.4, a triangularity of 0.45, $\beta_p=0.6$, $\beta_N=0.5$ and $q_{95}=4.5$.}

The information on the equilibrium has been extracted from the IMAS database~\cite{Imbeaux_NF2015}. The radial profiles of density and temperature are shown in Fig.~\ref{fig:profiles_nT_3.5s} (a) and (b) {with respect to the square root of the poloidal magnetic flux, $\sqrt{\psi}$.} In the absence of external torque, the plasma rotation is expected to be negligible, as inferred from other studies of tungsten transport in WEST \cite{Yang_NF2020,Maget_PPCF2023}. A tungsten concentration of $n_{ W}/n_{ D}=5.5\times10^{-5}$ is considered based on bolometer measurements. A nitrogen concentration of $n_{ N}/n_{ D}=2.5\%$ has been considered when nitrogen is included in the plasma. It corresponds to a concentration of $n_{ N}/n_{\rm e}=2.1\%$ of the total electron density. This concentration is typical of WEST and corresponds to a resistive $Z_{\rm eff}$ of $\sim2$. { In this manuscript, nitrogen impurities are typically in the banana-plateau regime and tungsten impurities are in the high-collisional Pfirsch-Schl\"uter regime.}

Electromagnetic effects are ignored for this plasma, because the beta parameter,
$\beta=2\mu_0 n_e T_e/B_{\rm T}^2$,
is low with an average value of $\beta\simeq10^{-4}$, see Fig.~\ref{fig:profiles_nT_3.5s} (c). The electrostatic model of XGC employed in this work is introduced in Sec.~\ref{sec:XGC}. 

The parameter $\rho^\star=\rho_{\rm L}/a$ is small with respect to typical turbulence simulations as the minor radius $a$, represents several hundreds thermal Larmor radius, $\rho_{\rm L}$, see subplot (d) that shows it inverse $1/\rho^\star$. For comparison, ITER plasmas have $\rho^\star\sim1/1000$. Nonlinear turbulent simulations, which are presented in this work, have thus required a high resolution grid (millimeter resolution). The most intense computations required for such multi-species high-resolution turbulence simulations have been performed on the Exascale supercomputer Frontier (OLCF). 

The turbulence is not always solved in XGC. For example, when studying collisional transport, turbulence is ignored and only the axisymmetric electric field is retained~\cite{Dominski2024}. In this case, a grid with a much coarser resolution can be used as the scale lengths of interest are the ion orbit widths and the inverse gradient lengths, which are significantly larger than the ion Larmor radius in this plasma. 

\begin{figure}
    \centering
    \includegraphics[width=8cm]{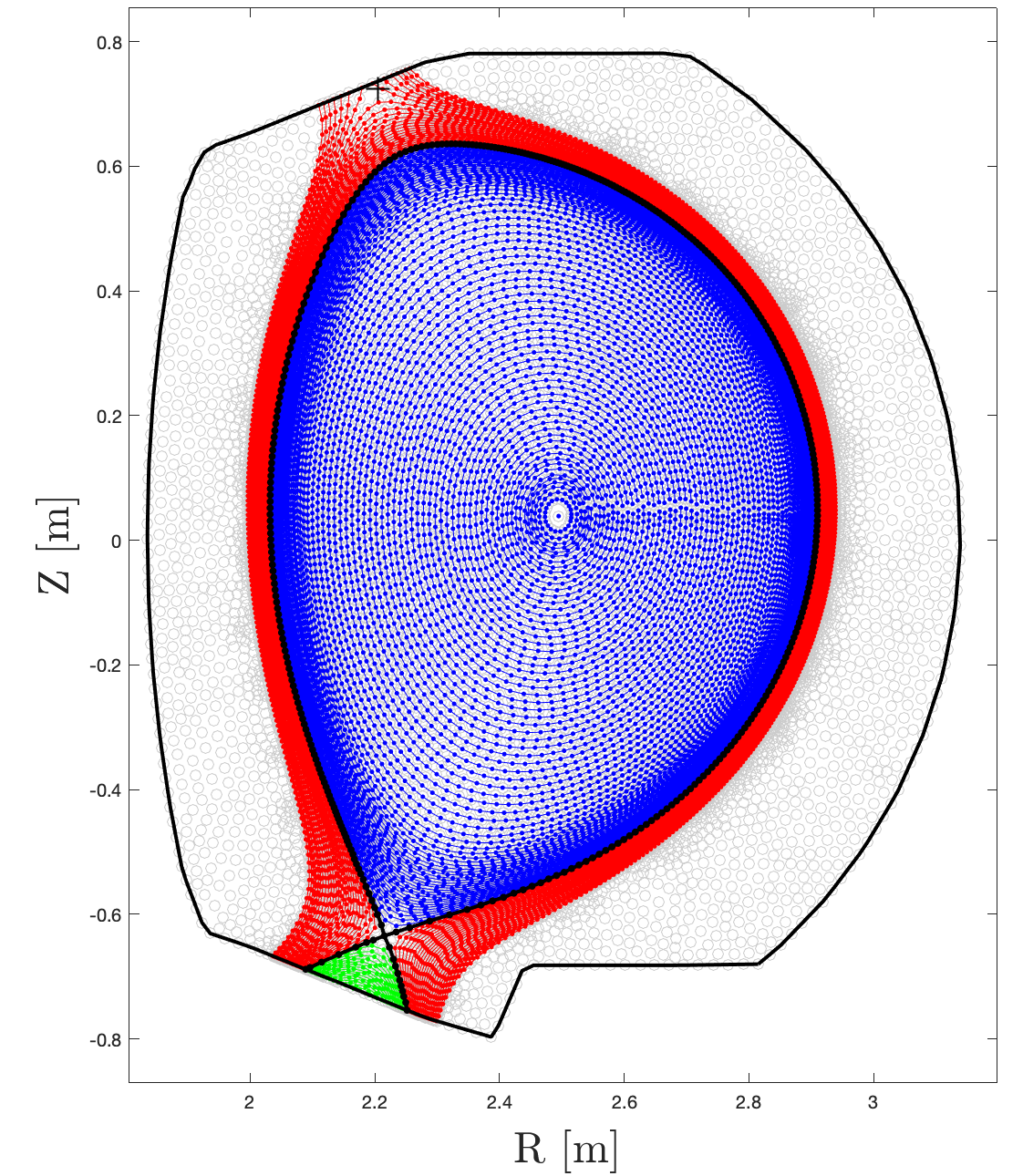}
    \caption{Example of an XGC grid used for collisional transport studies with the axisymmetric version of the code, namely ``XGCa''. The closed field line region is in blue, the scrape-off-layer region is in red, and the private region is in green. The separatrix and the wall are in black. The wall of WEST has been simplified to fit the magnetic data available on the IMAS database.}
    \label{fig:Surfaes_wall_and_grid_light_B}
\end{figure}

An example of coarse grid used to model the collisional transport is shown in Fig.~\ref{fig:Surfaes_wall_and_grid_light_B}. The region of closed field line is in blue, the scrape-off-layer is in red, the private region is in green, and the separatrix and wall are in black. The wall has been simplified to fit the magnetic data provided by the IMAS database. 

In this work, the typical grid used to simulate the collisional transport contains $\sim30$k vertices and needs a single {poloidal plane (at any toroidal position)} because only the $n=0$ mode is retained in these simulations, whereas the typical grid used to simulate the turbulence contains $700$k vertices and needs $32$ poloidal planes over half a torus. There is a factor $\sim750\times$ in between the number of vertices of these two grids. This ratio highlights the advantage of carrying out axisymmetric simulation of collisional transport when sufficient. For example, when preparing the simulation setup or when simply studying the collisional transport. 

\section{XGC electrostatic total-f model}
\label{sec:XGC}
\subsection{Gyrokinetic system of equations}
XGC is a multi-species total-f gyrokinetic particle in cell (PIC) code that can simulate the whole plasma volume including the scrape-off-layer and divertor regions~\cite{Ku16,Ku18,Hager2022}. The multi-species version of XGC has been described in Ref.~\cite{Dominski19a,Dominski2024}. Note that in the current work, we do not use bundles of tungsten but a single tungsten species, because we focus our analysis in regions where the charge state of tungsten does not vary too strongly. For example in the core the charge state of tungsten varies between $\sim 20$ and $\sim30$ and we use $Z=25$. 

In XGC, the particle distribution function is modeled with the total-f representation
\begin{equation}    
  f(t)=f_0(t)+{\delta f}(t)
\end{equation}
 where $f_0(t)$ is a slowly relaxing background and  $\delta f(t)$ is a perturbation modeled with a particle-in-cell (PIC) representation. {In this manuscript, $f_0$ is a local Maxwellian. The use of canonical Maxwellian background, recently implemented in XGC~\cite{Trivedi24}, is left for future studies.}
 
The total-f gyrokinetic equation reads
\begin{equation}
\frac{\partial \delta f}{\partial t}+\dot{\mathbf{X}}\cdot\frac{\partial \delta  f}{\partial {\mathbf{X}}}+\dot{v}_\parallel\frac{\partial \delta  f}{\partial v_\parallel}=-\frac{df_0}{dt}+\mathcal{C}[f],
\label{eq:GKE_df}
\end{equation}
where $df_0/dt$ is computed with the direct delta-f scheme~\cite{Dominski2024}. 

The equations of motion read
\begin{eqnarray}
\dot{\mathbf{X}}&=&\frac{1}{B_{\parallel}^\star}\left(v_\parallel \mathbf{B}^\star-\frac{\mu}{q}\nabla B\times \mathbf{b}-
\nabla\langle\phi\rangle\times\mathbf{b}\right)\\
\dot v_\parallel &=&-\frac{\dot{\mathbf{X}}}{mv_\|}\cdot\left(\mu\nabla B+q\nabla\langle\phi\rangle\right),
\label{eq:GKE}
\end{eqnarray}
with $\langle\phi\rangle$ the gyroaveraged electrostatic potential, $v_\|$ the parallel velocity, $\mu$ the magnetic moment, $\mathbf{X}$ the gyrocenter, $\mathbf{b}=\mathbf{B}/B$, and $\mathbf{B}^\star=\mathbf{B}+\frac{mv_\|}{q}\nabla\times\mathbf{b}$.

The electrostatic field is solved with the quasi-neutrality equation (here expressed in the long wavelength approximation)
\begin{equation}
    -\nabla_\perp\cdot\frac{ \sum_an_am_a}{eB^2}\nabla_\perp\phi +\frac{e \tilde\phi}{T_e}=-n_e^{\rm n.a.}+\sum_{a}Z_a\bar{n}_a,
    \label{eq:poisson}
\end{equation}
where $\bar{n}_a$ is the gyro-averaged density of each ion species $a$. Note that in the electrostatic version of XGC, the electron density $n_e$ is split into an adiabatic response, $\frac{e \tilde\phi}{T_e}$, on the left hand side of Eq.~(\ref{eq:poisson}) and a non-adiabatic response, $n_e^{\rm n.a.}$, on the right-hand-side of Eq.~(\ref{eq:poisson}). The axisymmetric version of XGC solves the axisymmetric version of the quasi-neutrality equation, see Eq.~(1) in Ref.~\cite{Dominski2024}.

The collision operator $\mathcal{C}$ is a nonlinear Fokker-Planck collision operator, see Ref.~\cite{Eisung14,Hager16} and all inter-species collisions are accounted for,
\begin{equation}
    \mathcal{C}[f_a]=\sum_b\mathcal{C}[f_a,f_b].
\end{equation}

Finally, the radial particle and kinetic energy fluxes are respectively defined by
\begin{equation}
\Gamma=\left\langle \int dv_\|d\mu \frac{B_\|^\star}{m}\ \nabla\psi\cdot \dot{\bf{X}} \  f\right\rangle_{\mathcal{S}}
\end{equation}
and
\begin{equation}
Q_=\left\langle \int dv_\|d\mu \frac{B_\|^\star}{m}\ \nabla\psi\cdot \dot{\bf{X}} \left( \frac{mv_\|^2}{2}+\mu B\right)\  f\right\rangle_{\mathcal{S}},
\end{equation}
where 
$\langle\ \rangle_{\mathcal{S}}$ is a surface average operation. The turbulent or ``E$\times$B'' fluxes are obtained by replacing $\dot{\bf{X}}$ by $\mathbf{v}_{E}=\frac{1}{B_\|^\star}\frac{\bf{E}\times\bf{B}}{B}$ and $f$ by $\delta f$.

{
The polarization flux due to the relaxation of the radial electric field,
\begin{equation}
    \Gamma_{\rm pol}=\left\langle \int dv_\|d\mu \frac{B_\|^\star}{m}\  \frac{1}{\Omega B}\frac{dE_\psi}{dt} \  \delta f \right\rangle_{\mathcal{S}},
    \label{eq:poldrift}
\end{equation}
is taken into account when estimating the peaking factor, in Sec.~\ref{sec:col}. 
}

\subsection{Larmor radius and cyclotron frequency of tungsten}
\label{sec:larmor}
In most tokamaks, the tungsten ions at thermal equilibrium have a smaller Larmor radius than deuterium as long as their charge number $Z=q/e$ is larger than
\begin{equation}
    Z_c=\sqrt{184/2}\simeq 9.6,
\end{equation}
which is typical for most of the plasma volume. Only in the scrape-off-layer region, the electron temperature of about $30-100$eV is small enough so that the tungsten charge number can be smaller than $9$. Nonetheless, even in this edge region, the tungsten Larmor radius remains close to the one of deuterium. For instance with $T_e\simeq10$eV the tungsten Larmor radius is only three times larger than the deuterium one.

These estimates are based on the equilibrium tungsten charge $\langle Z\rangle$ for which the ionization and recombination interactions are at equilibrium while using the rates from the ADAS database~\cite{Putterich_2010}. Note that a rough estimate of the tungsten Larmor radius is
$\langle Z\rangle\propto\sqrt{T_e}$ where $T_e$ is the electron temperature expressed in units of eV. This approximation provides an estimate of the tungsten Larmor radius
\begin{equation}
\rho_{W,th}\simeq\sqrt{\frac{T_w}{T_{e}}} \frac{1.4\times10^{-3}}{B},
\end{equation}
where $B$ is the magnetic field strength in units of Tesla and $T_{W}$ is the tungsten temperature. 

Finally, the cyclotron frequency of tungsten, which is proportional to $q/m$, can become significantly slower than the one of deuterium, even when their Larmor radius are similar. Indeed, for $Z=10$ (scrape-off-layer) the Larmor radii are equals but the tungsten cyclotron frequency is one order of magnitude smaller than the one of deuterium, as
\begin{equation}
\frac{\Omega_{W}}{\Omega_D}=\frac{Z}{92}.
\end{equation}
Nonetheless, given that typical wave frequencies are in general proportional to the species thermal velocity or to the sound speed, these frequencies of interest will also slow down in the edge proportionally to $\sqrt{T}$. Therefore, given that $Z\simeq\sqrt{T_{e}}$ for tungsten, the gyrokinetic-ordering still holds in the edge region, at leading order and far enough from the wall.

\section{Comparison between {an XGC edge simulation} and WEST data}
\label{sec:validation}
Heat deposition on target is important for modeling the tungsten sputtering from the wall. Therefore, we carried-out two exercises in the edge region of this WEST plasma. These exercises concern the energy flux level at separatrix and the heat load width at {mid-plane} and on target.  This effort provides confidence in the capability of the current simulation setup to model WEST turbulent transport. A dedicated study and validation of WEST turbulence plasma in the scrape-off-layer (SOL) is left for a future work when addressing the sputtering of tungsten ions from the wall.

{In this section we carry out a short simulation in which the edge transport reaches a  saturated level consistent with the local edge gradients. The accuracy of this level of transport depends on the initial profiles as these profiles did not have the time to relax consistently with respect to the core transport. Therefore, prior to estimating the heat load width, we check that the level of transport modeled in the edge is consistent with the one measured in the experiment at the separatrix.}

\begin{figure}
    \centering
    \includegraphics[width=8cm]{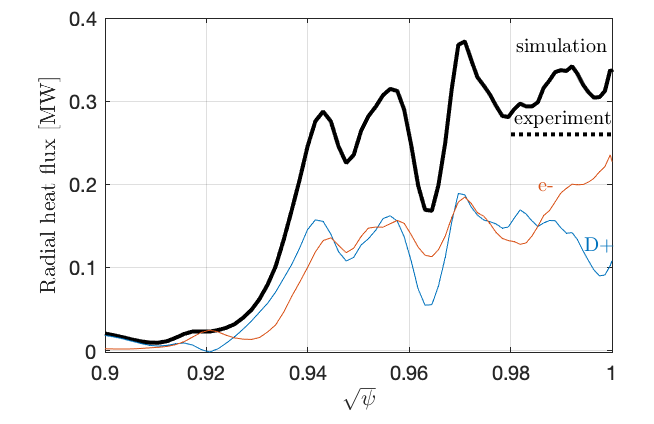}
    \caption{Comparison between XGC energy flux level in the pedestal at $t=0.295ms$ (full black line) with the power measured at the separatrix (horizontal dotted line) in MW. The electron and deuterium energy fluxes are shown in red and blue, respectively. They have similar levels, as expected.}
    \label{fig:heat_flux_edge_XGC1_t=0.295ms}
\end{figure}

The level of the energy flux measured at the separatrix in the XGC simulation, see the black solid line in Fig.~\ref{fig:heat_flux_edge_XGC1_t=0.295ms}, is in relatively good agreement with the measurement of the net power crossing the separatrix of WEST, see the black dotted line in the same figure. This net power is the difference between the injected power (Ohmic, ICRH, LH) and the radiated power. The flux computed in XGC is the sum of the electron, deuterium, and tungsten contributions measured in a nonlinear total-f turbulent simulation where all species are treated kinetically and collide together, as described in the previous section. The electron and deuterium dominate this heat flux as tungsten is a trace. The level of agreement reached is appreciable given the uncertainty on the profiles of density and temperature~\cite{Maget_PPCF2023,Artaud_2018}. This result thus provides confidence in the capability of our setup to model a transport relevant to this WEST discharge. Note that the snapshot of the profile of energy flux is measured when the turbulence is saturated in the edge region. At this time, the turbulence is still growing in the core, explaining why the flux is not yet developed before $\psi=0.85$.

\begin{figure}
    \centering
    \includegraphics[width=8cm]{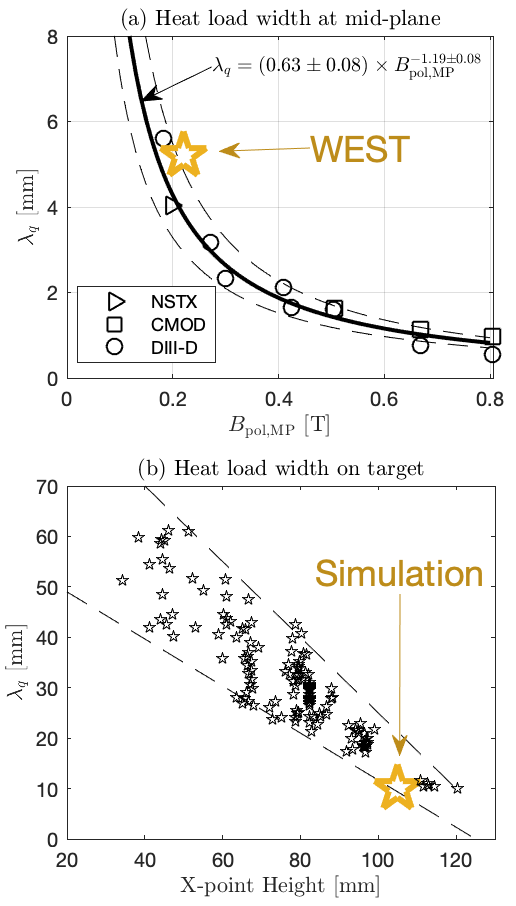}
    \caption{Heat load width at (a) mid-plane and (b) on target. (a) Heat load width at mid radius measured in our WEST simulation (yellow star) is compared with the Eich formula~\cite{Eich_2013} (solid black line) and with previous XGC estimates (black markers) published in Fig.~11 of Ref.~\cite{Chang17}. 
    (b) Heat load width on target (yellow star) is compared to experimental measurements (black triangle markers) published in Fig.~6 of Ref.~\cite{Gaspar21}.
}
    \label{fig:validation_heatloadwidth}
\end{figure}

The heat load width deposition at mid-radius is shown in Fig.~\ref{fig:validation_heatloadwidth} (a) where it is compared to the Eich formula~\cite{Eich_2013} (black curves) and to former XGC simulations~\cite{Chang17} (black markers). The width measured in our XGC simulation of WEST is indicated with a yellow star. It is worth pointing that the Eich formula has been established for low recycling H-mode discharges, which is not the regime of the plasma considered in this work. Nonetheless, the XGC estimate only lies 50\% above the Eich prediction, { which could be due to the fact that we study an L-mode plasma~\cite{Chang17}}. Many other factors could also explain this difference, including the fact that our plasma is in the ohmic phase, the density and temperature profiles have significant uncertainty, and various impurities are present in the edge. Such detailed study is left for future considerations.

The heat load width is also measured on target and compared to WEST results~\cite{Gaspar21} in Fig.~\ref{fig:validation_heatloadwidth}~(b). Again, the width measured in the XGC simulation of WEST is indicated with a yellow star. The black triangle markers represent measurements made on many WEST plasmas and extracted from Ref.~\cite{Gaspar21}. These results from Gaspar et al. are populated from plasmas in different regimes than the one we considered. The interesting aspect is the trend of the heat load width on target that decreases when the X-point height increases. As shown in this figure, the setup we considered is close to the largest X-point heights achieved in WEST, explaining the narrow heat load width we measure on target, see the yellow star.

\section{Collisional transport and peaking factor of tungsten in presence of nitrogen impurities}
\label{sec:col}

In the following section, we analyze the influence of nitrogen impurities on the tungsten peaking factor, asymmetries, and friction force. 

\subsection{Tungsten ``neoclassical'' peaking}
\begin{figure}
    \centering
    \includegraphics[width=8cm]{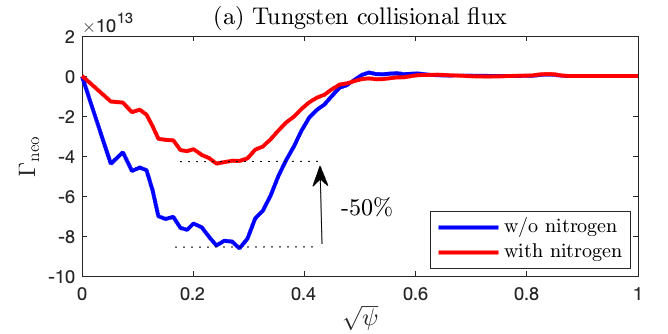}
    
    \includegraphics[width=8cm]{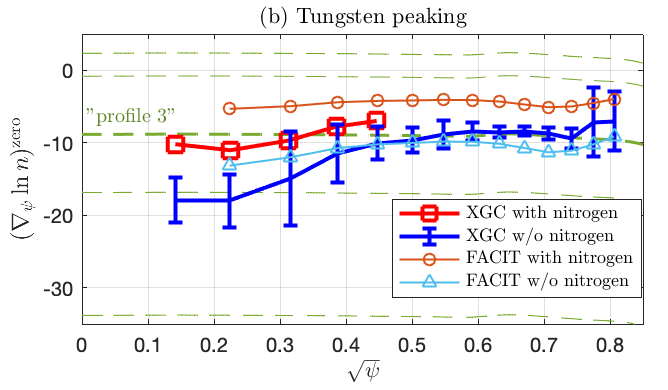}
    
    \caption{Collisional transport of tungsten particles in a deuterium plasma with $2.5\%$ of nitrogen impurities (red) and without nitrogen impurities (blue). {(a) Tungsten particle flux obtained when using the particular tungsten density profile ``profile 3''.} (b) Peaking factor of tungsten interpolated between different XGC simulations which were carried-out with different profiles of tungsten density (shown with green dashed lines). The peaking factor is verified with the FACIT model. {Fluxes are averaged over a short time window ($\lesssim 0.5$ms) at the end of the simulation.}}
    \label{fig:gradnln_wrt_Nprofiles_light_bis}
\end{figure}
The collisional transport of tungsten particles measured { at the end of} an axisymmetric XGC simulation is shown in Fig.~\ref{fig:gradnln_wrt_Nprofiles_light_bis} for a logarithmic gradient close to $\nabla\ln n_w = -9$ which is the third green dashed line labeled ``profile 3'' in subplot (b). It is found that nitrogen significantly affects the level of tungsten transport. In this case, the tungsten flux decreases by $50\%$ when adding $2.5\%$ of nitrogen, see the red curve in subplot (a). {Note that after $\sqrt\psi>0.5$, the particle flux is almost zero for this arbitrarily chosen ``profile 3'', because this profile is close to the peaking factor after $\sqrt\psi>0.5$ as we will now discuss.}

To provide a more representative influence of nitrogen on the tungsten collisional transport, we estimate the peaking factor from many XGC simulations carried out with different profiles of tungsten density, see Fig.~\ref{fig:gradnln_wrt_Nprofiles_light_bis} (b). At each radial location, we thus have the relation between the particle flux and the density gradient. We can then interpolate the steady-state gradient, $(\nabla\ln{n}_{W})^{\rm zero}$, giving a zero flux of particle, \textit{i.e.},
\begin{equation}
\Gamma^{\rm col}[(\nabla\ln{n}_{W})^{\rm zero}]=0.
\end{equation}
This estimation is made independently at each radial surface, by close interpolation~\footnote{First, we select two simulations labeled a and b such that $\Gamma_{\rm a}^{\rm col}$ and $\Gamma_{\rm b}^{\rm col}$ have opposite signs at this surface. Then, noting the gradient $y_i=(\nabla\ln{n}_{W})_{i}$ and its associated flux $x_i=\Gamma_{i}^{\rm col}/n_i$ , the steady-state gradient is computed with
\[
(\nabla\ln{n}_{W})^{\rm zero}=y_{\rm a}-\frac{y_{\rm b}-y_{\rm a}}{x_{\rm b}-x_{\rm a}}x_{\rm a}.
\]
}. The different profiles of tungsten density used to estimate the fluxes of particles are shown with horizontal green dashed lines in (b). The one with a thick green dashed line and labeled ``profile 3'' has a gradient about $\nabla\ln{n_{W}}\simeq9$ and will be used in complementary analysis below. 

A reduction of the neoclassical peaking factor of tungsten is measured in presence of nitrogen near the magnetic axis, see Fig.~\ref{fig:gradnln_wrt_Nprofiles_light_bis} (b).  Note that no XGC result is shown beyond $\sqrt\psi\gtrsim0.4$ for the results with nitrogen, because large poloidal oscillations of the tungsten density fluctuate and interfere with the convergence of the flux at mid-radius, as we will discuss in the following Sec.~\ref{sec:slowrelax}. 

The steady-state tungsten gradient is compared  with the results obtained with the FACIT code~\cite{Maget_PPCF2020b,Maget_PPCF2022,Fajardo_PPCF2022}. For using FACIT in a multi-ions species plasma, we assume that the tungsten flux resulting from the friction with deuterium and nitrogen is approximately the sum of the separate fluxes while enforcing electroneutrality, 
\begin{equation}
\Gamma_{W\rightarrow (D+N)}\simeq \Gamma_{W\rightarrow D} + \Gamma_{W\rightarrow N}.
\end{equation}
This assumption is justified in the limit where inter-species collisions are weakly dependent on the 3rd one, and when the coupling between horizontal and vertical asymmetries is negligible, which is the case at vanishing tungsten flux \cite{Maget_PPCF2020b}. In the present case without external drives (rotation or electrostatic potential asymmetry), the tungsten poloidal asymmetry at vanishing flux is expected to be zero \cite{Maget_PPCF2020}. 

There is a particularly good agreement between the two codes at mid radius in simulations without nitrogen (blue). XGC predicts a larger peaking factor than FACIT near axis, but the agreement is still good as FACIT results fall within XGC error bar. This error bar is estimated by accounting for the standard deviation of the time averaged flux when interpolating for the zero flux of tungsten particles. FACIT recovers the reduction of the neoclassical tungsten peaking in presence of nitrogen impurities. 

We recall that XGC is not a neoclassical code but a gyrokinetic code, therefore the collisional transport is not solved with respect to the neoclassical ordering but with respect to the gyrokinetic ordering, which contains higher order terms, as discussed in Ref.~\cite{Dominski2024}. Collisional transport of XGC agrees with the one of neoclassical codes when these ordering are close enough, as shown in Ref.~\cite{Hager16,Dominski19a}. Differences can be explained by this ordering difference. 

\begin{figure}
    \centering
    \includegraphics[width=7.5cm]{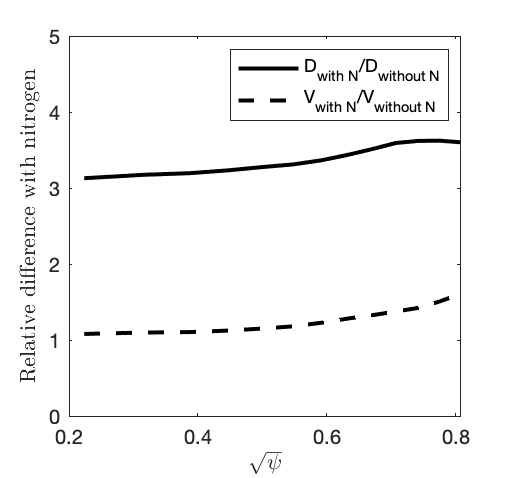}
    \caption{Relative change of $D$ and $V$ when including nitrogen, where $\Gamma_{\rm neo}=-D_W \nabla n_W + n_W V_W$. The change in the diffusivity explains the change in the peaking factor $\nabla \ln{n_W}=V_W/D_W$.}
    \label{fig:D_and_V_with_without_N}
\end{figure}

{Finally, the reduction of the peaking factor, $(\nabla\ln{n_W})^{\rm zero}=V/D$, when nitrogen is added to the ion mix originates from an increase of the diffusion coefficient $D$ rather than a decrease of the pinch velocity $V$, as shown in Fig.~\ref{fig:D_and_V_with_without_N} presenting FACIT calculations~\cite{Maget2024}. Indeed, the diffusion coefficient scales as $D\propto ZA^{1/2}$ making it sensitive to the main ion charge $Z$ and atomic number $A$, whereas the pinch velocity is mainly determined by the tungsten charge in this Pfirsch-Schl\"uter collisional regime. The diffusion $D$ and velocity $V$ relate to the particle flux via the relation
\begin{equation}
    \Gamma_{\rm neo}=-D \nabla n_W + n_W V_W.
\end{equation}
}

\subsection{Oscillation of the tungsten asymmetry in presence of nitrogen impurities}
\label{sec:slowrelax}
\begin{figure}
    \centering
    \includegraphics[width=8.5cm]{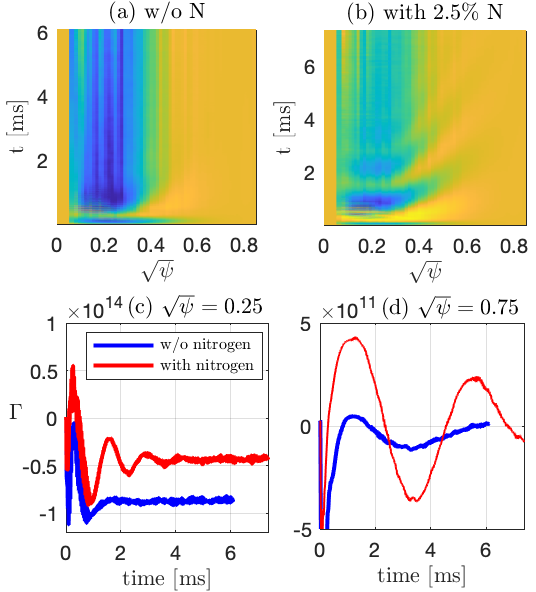}
    \caption{Radial particle flux of tungsten. Time evolution of the radial profile (a) without and (b) with nitrogen. Time evolution of the particle flux at (c) $\sqrt{\psi}=0.25$ and (d) $\sqrt{\psi}=0.75$, without nitrogen (blue) and with nitrogen (red).}
    \label{fig:neoclassic_combin_DN}
\end{figure}
Slow oscillations, $\omega_{W}=750$Hz, are observed on the particle flux of tungsten. These oscillations are due to the transient between our initial conditions and the convergence of the tungsten asymmetries. In Fig.~\ref{fig:neoclassic_combin_DN} (b), they are more visible near the axis at $\sqrt{\psi}\simeq0.25$ where the particle fluxes are larger. Near the axis, which is the region of interest, these slow oscillations are damped in about 1.5ms in the plasma without nitrogen and in about 5ms in the plasma with nitrogen, see subplot (c). The tungsten density profile used in these simulations has a logarithmic gradient equal to $\nabla\ln{n}_{W}\simeq-9$ and corresponds to the ``profile 3'' in Fig.~\ref{fig:gradnln_wrt_Nprofiles_light_bis}(b). 

These oscillations are also visible on the tungsten density asymmetry. Such asymmetry can be seen as an $m=1$ perturbation written
\begin{equation}
n_{W}/\left\langle n_W\right\rangle_{\mathcal{S}}=1+\delta\cos\theta+\Delta\sin\theta,
\end{equation}
where $\delta$ is the in-out asymmetry, $\Delta$ is the up-down asymmetry, and $\theta$ is the poloidal angle equal to zero at the outer mid-plane.

\begin{figure}
    \centering
    \includegraphics[width=8cm]{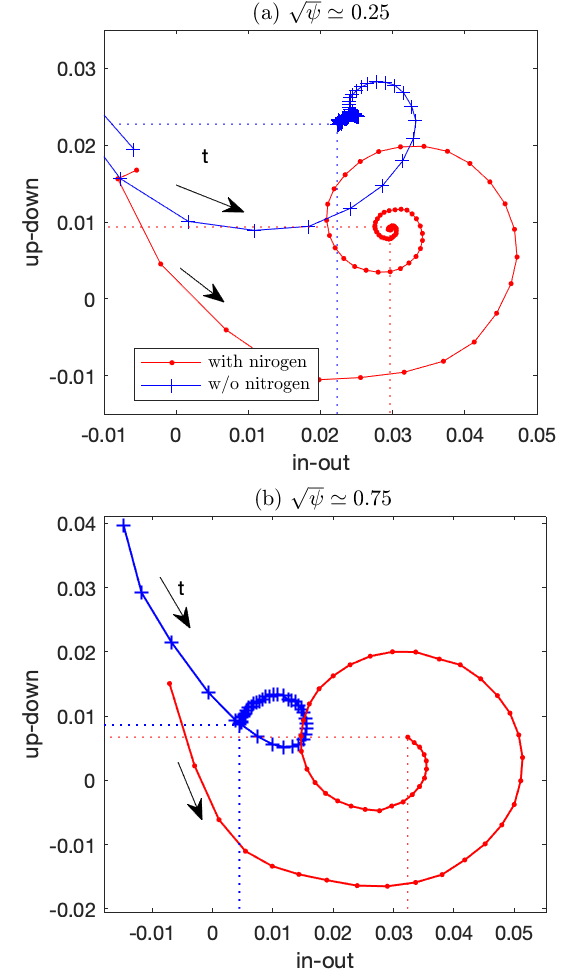}
    \caption{Evolution of the tungsten density asymmetry in simulations carried-out with (red) and without (blue) nitrogen, at radial positions (a) $\sqrt\psi\simeq0.25$ and (b) $\sqrt\psi\simeq0.75$. The black arrow indicates the direction of the time evolution (counter clock-wise in this case). The red asterisk represents the nitrogen final asymmetry which converges pretty fast, \textit{i.e.}, in about 1.5ms.  
    }
    \label{fig:phase_asymmetry}
\end{figure}
The evolution of these tungsten asymmetries is shown in Fig.~\ref{fig:phase_asymmetry}, where we plot the evolution of the up-down asymmetry with respect to the in-out asymmetry. These orbits in $(\delta,\Delta)$ space are shown (a) near the axis for $\sqrt{\psi}\simeq0.25$ and (b) closer to the edge near $\sqrt{\psi}=0.75$. The arrow of time is indicated in black. These curves rotate counter-clockwise and are reminiscent of fluid tungsten transport simulations carried out with the XTOR-2F code \cite{Maget_PPCF2020}.

In presence of nitrogen (red curves) the asymmetries take $3\times$ more time to converge than without nitrogen (blue curves). It matches the $3\times$ smaller damping rate of oscillations seen on the particle flux. It is worth pointing that the parallel flow of tungsten is also decreased by a factor $3\times$ in presence of nitrogen, see next section. {In other words, nitrogen impurities reduce the tungsten parallel velocity and thus slow down the convergence of the tungsten asymmetry by slowing down its dynamical evolution.}

{Given the slow down of the tungsten parallel dynamics in presence of nitrogen,} the red orbits describe several revolutions before landing on their respective attractor. For the case near $\sqrt\psi=0.75$ in (b), the simulation with nitrogen is not yet converged after almost $7$ ms and 25k steps of simulations. In comparison, the $(\delta,\Delta)$ orbits of the simulations without nitrogen (blue) converge much faster, which is consistent with the faster convergence of the particle flux observed in Fig.~\ref{fig:neoclassic_combin_DN}.

\subsection{Reduction of the tungsten friction force and up-down asymmetries in presence of nitrogen impurities}

In presence of nitrogen impurities, the up-down asymmetry of tungsten density measured at $\sqrt\psi\simeq0.25$ in Fig.~\ref{fig:phase_asymmetry}~(a) is reduced by $\sim60\%$, which is similar to the reduction of the particle flux observed in Fig.~\ref{fig:neoclassic_combin_DN}~(b) and (c). Indeed both the particle flux and the up-down asymmetry are the result of the effect of the friction force which is also estimated to be reduced by $\sim60\%$.

\begin{figure}
    \centering
    \includegraphics[width=8cm]{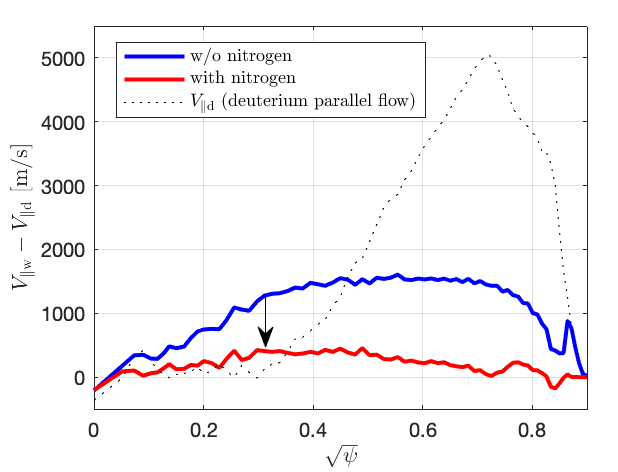}
    \caption{Difference between the tungsten parallel flow and the main ion (deuterium) parallel flow, in simulations carried-out with (red) and without (blue) nitrogen impurities. The black dashed line represents the main ion flow which does not significantly vary in presence of nitrogen ions. When using nitrogen impurities, the concentration is $n_{ N}/n_{ D}=2.5\%$.}
    \label{fig:N3T2_with_and_without_N_V_parallel}
\end{figure}

The parallel friction force of tungsten colliding on deuterium can be estimated~\cite{Esteve2018} with
\begin{equation}
    R_{\|W}=-m_{W}\,n_{W}\,\nu_{W D}\left(V_{\|W}-V_{\| D}-\frac{3}{5}\frac{q_{\| D}}{n_{ D}T_{ D}}\right),
\end{equation}
where $V_\|$ is the parallel flow of particles, $q_\|$ is the parallel flow of heat, and $\nu_{WD}$ is the collisionality of tungsten ions on deuterium ions defined in~\cite{Dominski19a}. Near axis, $\sqrt\psi<0.25$, the parallel flow of deuterium is nearly zero, see the black dotted line in Fig.~\ref{fig:N3T2_with_and_without_N_V_parallel}, and so is its parallel heat flow, $q_{\|}$. The remaining drive for the parallel friction force is the parallel flow of tungsten, which is reduced by $66\%$ in presence of nitrogen, see blue and red curves in Fig.~\ref{fig:N3T2_with_and_without_N_V_parallel}. Note that the parallel flow of deuterium is not affected by the presence of nitrogen, whereas the parallel flow of tungsten becomes equal to the one of nitrogen when nitrogen impurities are present. 

\section{Influence of turbulence on the collisional peaking of tungsten, in presence of nitrogen}
\label{sec:turbulence}
\begin{figure}
    \centering
    \includegraphics[width=8cm]{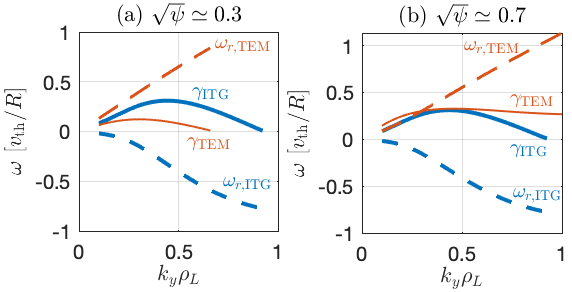}
    
    \caption{Local analysis of micro-instability at $\sqrt{\psi}\simeq0.3$ and $0.7$ in subplot (a) and (b), respectively. ITG branch is blue and TEM branch is red. Growth rates are in full lines and real frequencies are in dashed lines.}
    \label{fig:local_analysis}
\end{figure}

The core of this ohmic plasma is in the ion temperature gradient (ITG) micro-turbulent regime{, driven by the deuterium ions}. Trapped electron modes (TEMs) are secondary and become significant near the edge $\sqrt{\psi}\gtrsim0.7$. This has been identified with XGC and verified with a local dispersion relation, see Fig.~\ref{fig:local_analysis} showing the ITG and TEM branches computed at the radial positions $\sqrt{\psi}=0.3$ and $0.7$. The local dispersion relation is described in Ref.~\cite{Dominski15}.

\begin{figure}
    \centering
    \includegraphics[width=8cm]{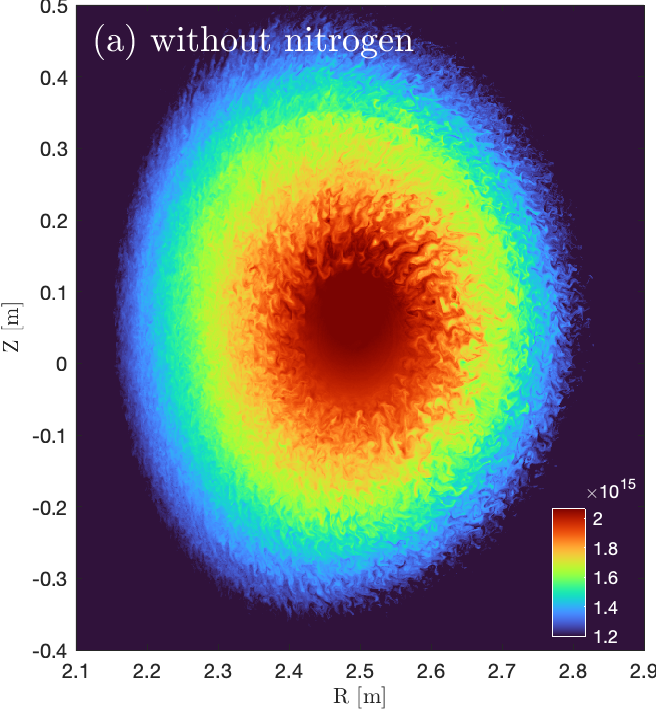}
    
    \includegraphics[width=8cm]{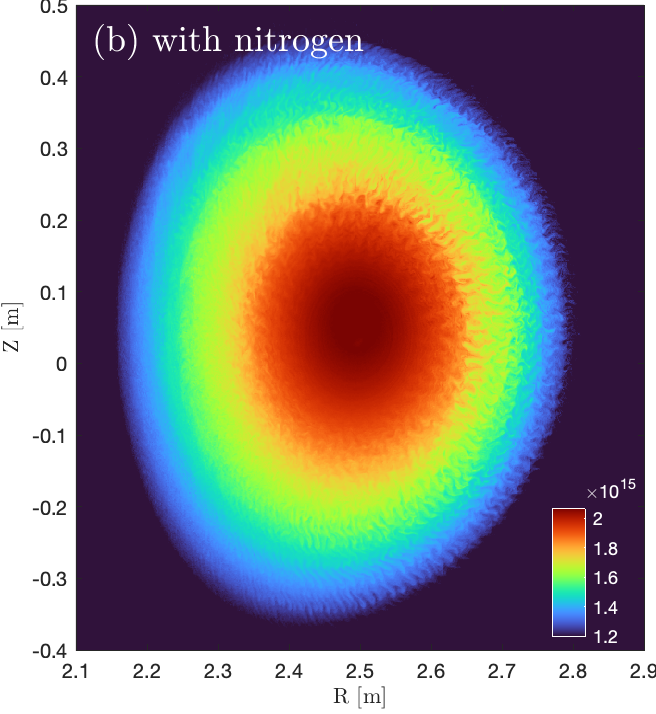}
    
    \caption{Tungsten turbulent density (a) without and (b) with nitrogen impurities in the plasma. Electrons are kinetic and full-f collisions between all kinetic species are activated.}
    \label{fig:XGC1_tungsten_density_with_without_N}
\end{figure}

\begin{figure}
    \centering
    \includegraphics[width=8.5cm]{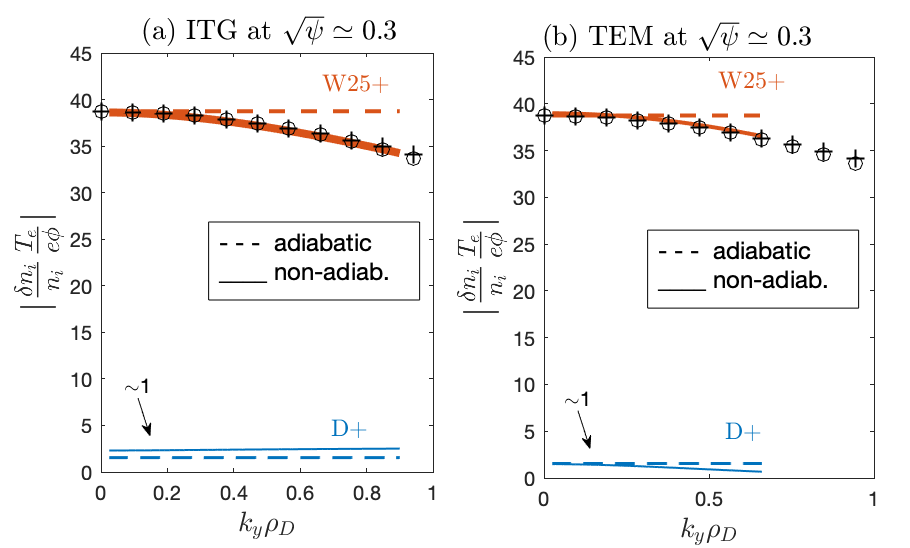}
    \caption{Ion perturbation relative to the background for deuterium (blue) and tungsten (red). These responses are normalized to the relative perturbation of adiabatic electrons $e\phi/T_e$. Black crosses and circles are estimated with the absolute value of Eq.~(\ref{eq:na_tungstena}) and Eq.~(\ref{eq:na_tungstenb}), respectively. Not reported, but we also verified that forcing $J_0^2=1$ for tungsten in the code remove this reduction from the non-adiabatic response.
    }
    \label{fig:ions_responses}
\end{figure}

{
\subsection{Tungsten turbulence amplitude}
Turbulent eddies and streamers have a large amplitude of $\sim30\%$ compared to the background density, see  Fig.~\ref{fig:XGC1_tungsten_density_with_without_N} (a). Indeed, even if tungsten ions are essentially a trace impurity with $\delta n_W\ll\delta n_e$, the relative perturbation $\delta n_W/n_W$ can be large.

This large perturbation is mainly the result of the tungsten evolving passively in the $E\times B$ turbulence. Even though the adiabatic response of tungsten can be large~\cite{Esteve2018} it is essentially canceled by its non-adiabatic response, see Fig~\ref{fig:ions_responses} where the deuterium and tungsten responses are estimated from the local relation Eq.(22) of Ref.~\cite{Dominski15},
\begin{equation}
    \frac{\delta n_j}{n_{0j}}=-\frac{eZ_j \phi}{T_j}\int d^3v\left[1-J_0^2\frac{\omega-\omega^\star}{\omega-k_\|v_\|-\omega_{dj}}\right]\frac{f_{0j}}{n_{0j}}.
    \label{eq:ionresponses}
\end{equation}
The term proportional to 1 in the square bracket is the adiabatic response and the term proportional to $J_0^2$ is the non-adiabatic response, where $J_0=J_0(k \rho)$ is the Bessel function of the first kind, $\omega$ is the mode frequency, $\omega^\star$ is the diamagnetic frequency, and $\omega_d$ is the drift wave frequency. 

Tungsten thus primarily responds passively to the $E\times B$ nonlinear turbulence driven by deuterium ITG, {\color{red}together with a mild linear} response at short scales where the adiabatic response is not fully canceled by the non-adiabatic response. This reduction of the non-adiabatic tungsten response is due to finite Larmor radius (FLR) effects, such that
\begin{eqnarray}
\frac{\delta n_W^{\rm n.a.}}{n_W}&=&-\frac{eZ_W\phi}{T_W} e^{-k^2\rho^2_W}I_0(k^2\rho_W^2)\label{eq:na_tungstena}\\
&\simeq&-\frac{eZ_W\phi}{T_W} (1-k^2\rho^2_W),
\label{eq:na_tungstenb}
\end{eqnarray}
as illustrated in Fig.~\ref{fig:ions_responses} with black crosses for Eq.~(\ref{eq:na_tungstena}) and black circles for Eq.~(\ref{eq:na_tungstenb}). For tungsten $W^{25+}$, only the term $J_0$ is significant because the related small terms $\omega^\star\propto1/Z$, $\omega_d\propto1/Z$ and $k_\|v_\|\propto1/\sqrt{m}$ are vanishing.

In the (a) deuterium driven ITG regime, which is the main drive of micro-turbulence in the studied plasma, the deuterium's non-adiabatic response is $50\%$ higher than its adiabatic response, see blue curves. In comparison, in the TEM regime (b), the deuterium's non-adiabatic response is close but smaller than its adiabatic response. This reduction is explained by FLR effects and by the ratio of frequencies in Eq.~(\ref{eq:ionresponses}), which frequencies are small but non negligible for deuterium ions.
}


\subsection{Tungsten stabilization by nitrogen}
The tungsten turbulence in these total-f XGC simulations has a significantly weaker intensity in presence of nitrogen, see Fig.~\ref{fig:XGC1_tungsten_density_with_without_N} subplot (b). 
Indeed, adding nitrogen to the plasma has a stabilizing effect~\cite{Bonanomi2018,Maget_2022}. One cause of this stabilization is the dilution of the main ion relatively to the electron. For example, the associated increase of the adiabatic electron response with respect to the main ion perturbation has a stabilizing effect, in particular for ITG micro-instability. This has been observed for this plasma close to the axis with the local dispersion relation. 

\begin{figure}
    \centering
    \includegraphics[width=8cm]{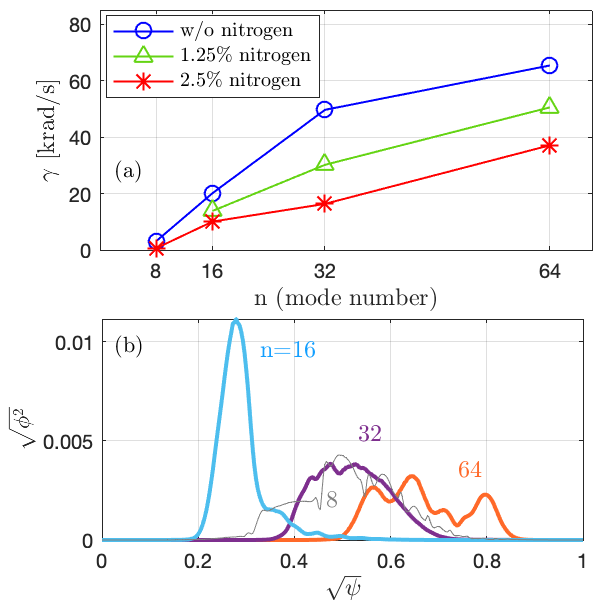}
    \caption{Linear stabilisation of the deuterium ITG by the presence of nitrogen. {(a) growth rates of different toroidal mode number, n. (b) radial profile of the different modes.} Compared to the nonlinear total-f simulation, here there is no tungsten, no collisions, no kinetic electrons, and we use a field-aligned filter and the reduced delta-f model instead of the total-f one.}
    \label{fig:profiles_linear_deltam=7n}
\end{figure}
The influence of the nitrogen ions on the ITG stability is quantified with a linear micro-instability analysis in Fig.~\ref{fig:profiles_linear_deltam=7n}. Results have been obtained from collision-less adiabatic-electron simulations with deuterium and nitrogen ions modeled gyrokinetically. The results show that the concentration of $n_{ N}/n_{ D}=2.5\%$ has a strong stabilizing effect on the linear ITG instability. Even half of this concentration, $n_{ N}/n_{ D}=1.25\%$, has a significant impact on the linear growth rates. In this linear analysis, we used a filter to only account for modes which are field aligned with the magnetic field. This filter is implemented in Fourier space and keeps only the poloidal mode numbers $m$ close to the resonant mode number $nq_s(\psi)$, such that $|nq-m|\le 7$. This filter is not used in the nonlinear simulations.

Given these first observations, it appears necessary to include nitrogen impurities for accurately modeling the physics of tungsten transport in this plasma. Indeed, as we will show in the following section, the predicted steady-state profile of tungsten density deviates significantly from the experimental one when ignoring the influence of nitrogen.

\subsection{Prediction of the tungsten peaking near axis and validation with a synthetic diagnostic}
\begin{figure}
    \centering
    \includegraphics[width=8cm]{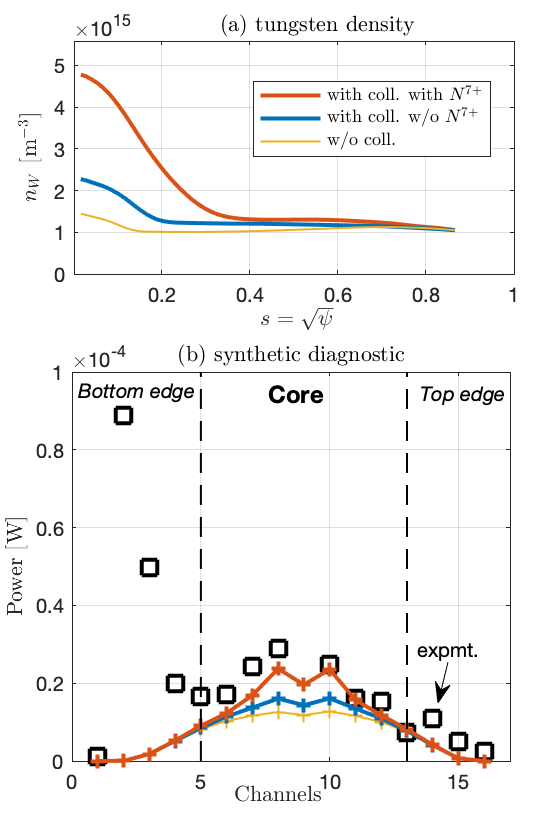}
    \caption{Prediction of (a) tungsten density peaking near axis in and (b) synthetic diagnostic of the radiation measured by the bolometers. {For the case w/o coll. (yellow), the presence of nitrogen was not found to have a significant influence. Only one simulation is thus reported here.}
    }
    \label{fig:all_profiles_light}
\end{figure}

The profile of tungsten in the core of WEST is predicted with XGC total-f turbulent simulations, see results in Fig.~\ref{fig:all_profiles_light}(a). This predicted density of tungsten is obtained by integrating over the steady-state density gradient (peaking factor),
\begin{equation}
    n_{W}(\sqrt{\psi})=n_{W}(0.85)+\int_{0.85}^{\sqrt{\psi}} ds\  (\nabla_s{n_{W})^{\rm zero}},
\end{equation}
where the density at the ``pedestal top'', $\sqrt{\psi}=0.85$, is taken from experimental data. 
For each setup (with or without nitrogen ; with or without collisions), the critical density gradient is estimated from two saturated turbulent simulations: one carried out with a fixed concentration of tungsten $n_{W}/n_{ D}=5\times10^{-5}$ and one obtained with an almost fixed density gradient $\nabla\ln{n_{W}}\simeq8$ at mid-radius and $\sim5$ near axis. 

This prediction is validated in Fig.~\ref{fig:all_profiles_light}(b) where the bolometer measurements are compared to a synthetic diagnostic~\cite{Devynck2021}. This synthetic diagnostic computes the power that would be measured by bolometers while using the profiles of tungsten calculated in (a). In this figure, three sets of turbulent simulations are compared. In yellow, the XGC prediction is obtained in turbulent simulations without collisions (nitrogen has a minor effect in these collision-less simulations), in blue we include collisions to the turbulent simulations, and in red we include collisions and nitrogen impurities into the turbulent simulations. The simulation labeled ``without nitrogen'' and ``with nitrogen'' in previous Fig.~\ref{fig:XGC1_tungsten_density_with_without_N} have been used to produce the results shown in blue and red in Fig.~\ref{fig:all_profiles_light}, respectively.

The different channels of bolometers indicate different lines of integration that have different orientations. The channels which are the most representative regarding our prediction are the ones going from 5 to 13, because these lines point towards the core where we model the plasma. The region around the axis is mainly measured by channels 8 and 10. In comparison, channels 1 to 4 point towards the bottom of the plasma aiming at the wall, the private region, and the X-point, where strong radiations occur as measured on channels 2 and 3. Similarly, channels 14 to 16 point towards the top of the plasma. 

\begin{figure}
    \centering
    \includegraphics[width=8cm]{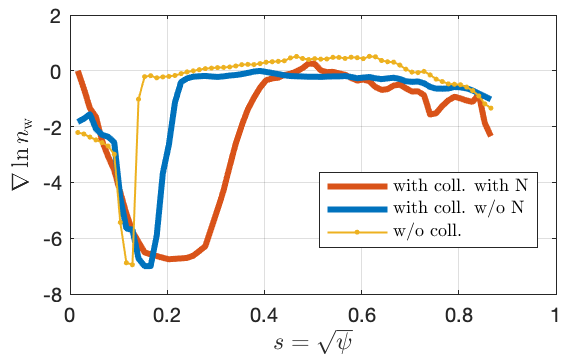}
    \caption{Peaking factors at vanishing particle flux of tungsten computed from three different total-f turbulent set-up: (red) with collisions and nitrogen impurities, (blue) with collisions but without nitrogen impurities, and (yellow) without collisions. {All electron and ion-species are modeled kinetically.} Due to the long time required to converge the tungsten collisional transport in presence of nitrogen impurities, the peaking factor near $\sqrt{\psi}\simeq0.2$ is estimated from collisional simulations.}
    \label{fig:prediction_peaking_and_profiles_agains_bolometrie}
\end{figure}

Based on this synthetic diagnostic, the most accurate prediction (red) is obtained when including turbulence, collisions, and nitrogen impurities. In which case more tungsten is predicted to peak near the axis, which explains the larger radiated power on channel 8 and 10. The reason why XGC predicts more tungsten peaking when nitrogen is accounted for is related to the fact that nitrogen stabilizes turbulence and shifts the boundary separating the region dominated by collisional transport near the axis with the region dominated by turbulent transport at mid-radius. This is shown by the peaking factor of these different plasmas in Fig.~\ref{fig:prediction_peaking_and_profiles_agains_bolometrie}. Therefore, even if nitrogen reduces the neoclassical peaking factor near the axis it also reduces the turbulence screening at mid-radius and thus broaden the region dominated by the strong neoclassical peaking. 

In this analysis we focus on the influence of nitrogen in the core, as we are not considering the impact of nitrogen on the tungsten sputtering from the wall. Indeed we keep a fixed density of tungsten near the pedestal top, $\sqrt{\psi}=0.85$. 

The influence of the heat source on the profiles is not accounted for. Indeed, if the plasma is adequately heated near the axis, the nitrogen stabilization of turbulence could increase the temperature gradient until the ITG is unstable again or until the collisional transport balances the input power. Conversely, if heating is inadequately applied near the axis, as seen during a radiative collapse~\cite{Ostuni2022}, a reduced temperature gradient will result in turbulence stabilization by nitrogen impurities thus leading to an increase of the tungsten peaking in the core. This increase of the tungsten density in the core would lead to an increase of the radiated power which would further cool down the electron temperature.

\subsection{Influence of electrons, collisions, and nitrogen on the fluxes}
\begin{figure}
    \centering
    \includegraphics[width=8cm]{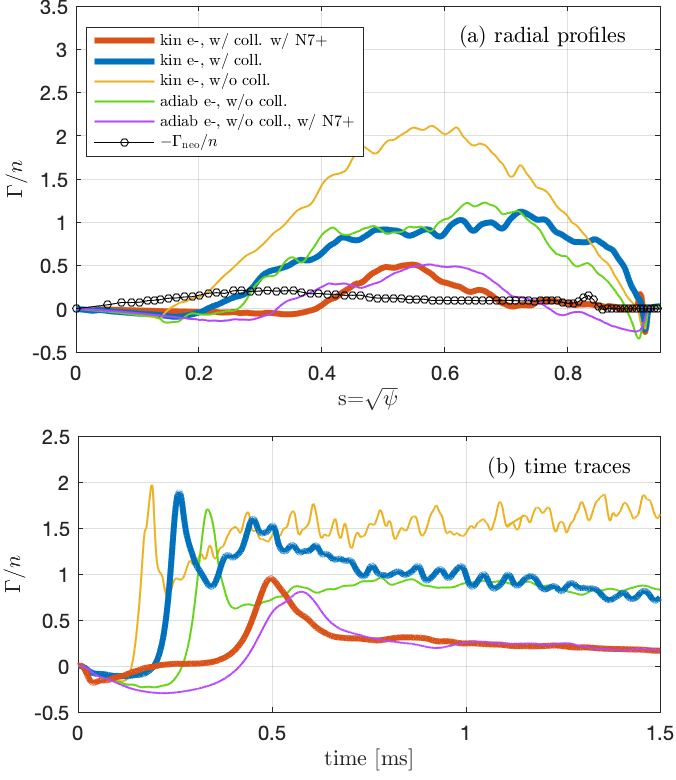}
    
    \caption{Tungsten turbulent particle flux normalized to its density, $\Gamma/n$. (a) Radial profiles averaged from $t=1$ to $1.5$ms. Color code in legend. (b) Time traces of the averaged flux.}
    \label{fig:oned_eDW2_flux_tungsten_reduced_vs_directdf_adiab_kine_manuscript_light}
\end{figure}

To illustrate the influence of the physics of electrons, collisions, and nitrogen on the tungsten fluxes, we carry-out and compare multiple XGC simulations by starting with a simple collision-less adiabatic electrons simulation without nitrogen (green) and gradually adding kinetic electrons (yellow), collisions (blue), and nitrogen (red), see Fig.~\ref{fig:oned_eDW2_flux_tungsten_reduced_vs_directdf_adiab_kine_manuscript_light}. In these simulations, a fixed concentration of tungsten, $n_{W}/n_{ D}=5\times10^{-5}$, is used. The color code used to identify the different simulations in the figures is summarized in Tab~\ref{tab:color_code}.
\begin{table}[]
    \centering
        \begin{tabular}{|c|c|c|c|}
        \hline
        \backslashbox{Color}{Model}  & electron & collisions & nitrogen \\
        \hline
        green  & adiabatic & w/o & w/o \\
        purple    & adiabatic   & w/o  & with  \\
        yellow & kinetic   & w/o & (w/o) \\
        blue   & kinetic   & with  & w/o \\
        red    & kinetic   & with  & with  \\
        \hline
    \end{tabular}
    \caption{Color code of nonlinear simulations}
    \label{tab:color_code}
\end{table}

Comparing the green and blue curves, it appears that the same radial profile of tungsten particle flux is obtained from the simplest collision-less simulation with adiabatic electrons (green) than in the complete simulation with kinetic electrons and collisions (blue). Similar observations are made for simulations carried out with nitrogen impurities and shown in purple (adiabatic electron w/o collisions) and red (kinetic electrons with collisions). { As we will further discuss at the end of this section, collisions are understood to reduce the destabilizing influence of the non-adiabatic electron response and associated turbulent fluxes.}

In this plasma dominated by instabilities driven by the main-ion and electron species, the tungsten turbulent particle transport, $\Gamma_{E\times B}$, is the result of the tungsten distribution function evolving ``passively'' on the $E\times B$ turbulence driven by the main ion. Therefore, the influence of nitrogen on tungsten turbulence is done through the stabilization of deuterium's ITG turbulence, which is felt by all species.

\begin{figure}
    \centering
    \includegraphics[width=8cm]{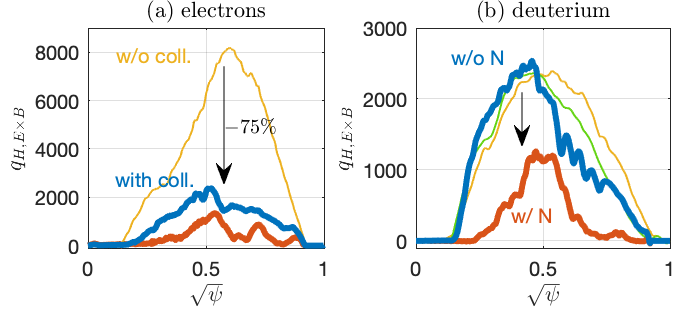}
    
    \caption{Turbulent $E\times B$ heat flux $q_{hE\times B}$ of the (a) electrons and (b) deuterium (main) ions. Same color code as in previous figures and summarized in Tab~\ref{tab:color_code}.}
    \label{fig:comparison_qHExB_total_with_without_coll3}
\end{figure}
This stabilizing influence of nitrogen on turbulence is highlighted by the turbulent heat flux,
\begin{equation}
q_{hE}=Q_{E}-\frac{5}{2}T\,\Gamma_{E},
\end{equation}
which is calculated for the (a) electrons and (b) deuterium in Fig.~\ref{fig:comparison_qHExB_total_with_without_coll3}. 
For each species, the $E\times B$ turbulent heat fluxes decreases when the simulation includes nitrogen (going from blue to red). This reveals the stabilization of ITG turbulence by nitrogen, which reduces the drive for the field perturbation $\phi$ and associated $E\times B$ turbulence. Note that in these turbulent simulations with collisions (blue and red), the ions and electrons have a similar level of  turbulent heat flux. 

Looking back at the influence of the non-adiabatic electron responses, it is interesting to point out that the simulation with kinetic electrons (yellow) has a larger particle flux than the simulation with adiabatic electrons (green) in Fig.~\ref{fig:oned_eDW2_flux_tungsten_reduced_vs_directdf_adiab_kine_manuscript_light}, but that these two simulations have the same level of ion heat flux in Fig.~\ref{fig:comparison_qHExB_total_with_without_coll3}(b). Indeed, the non-adiabatic response of electrons, 
\[
\delta h_e\simeq\delta f_{e}-f_{0}e\tilde{\phi}/T_e,
\]sustains a significant contribution to the turbulent particle flux~\cite{Dominski15,Dominski17} which in turn increases the other species particle flux. On the other hand,  when switching on collisions (blue curves), the non-adiabatic response of kinetic electrons is considerably reduced and the turbulent heat and particle fluxes go back to the levels measured in the adiabatic electron simulation (green). {Reduction of the non-adiabatic electron response and associated transport by collisions has been reported in Ref.~\cite{Mikkelsen08}.} The behavior of the kinetic electrons and collisions is the same in simulations including nitrogen impurities.

\section{Conclusion}
\label{sec:conclusion}

A study of the influence of nitrogen on the collisional and turbulent transport of tungsten in the core of a WEST plasma has been carried out with the XGC code. These total-f simulations include gyrokinetic ions of deuterium (D$^+$), tungsten (W$^{25+}$), and nitrogen (N$^{7+}$). Electrons are handled kinetically, and all full-f nonlinear inter-species collisions are accounted for. 

Nitrogen impurities are shown to reduce both the collisional peaking and collisional particle flux of tungsten in the core of this plasma{, up to $\sqrt{\psi}=0.85$}. {This reduction of tungsten peaking, which is cross-verified with FACIT, is due to the increase of the collisional diffusion of tungsten in presence of nitrogen, while the convective part of tungsten transport is barely changed.} On the other hand, the reduction of the tungsten flux and asymmetry in presence of nitrogen is related to a reduction of the parallel friction force caused by a reduction of the parallel flow, as the tungsten parallel flow reduces to the same level as the one of nitrogen. 

Nitrogen impurities are also shown to significantly reduce the tungsten turbulence by stabilizing the ion temperature gradient (ITG) mechanisms of the main ion (deuterium), which is the dominant source of turbulence in the core of this plasma. In turn, the turbulence screening of tungsten at mid-radius is considerably reduced and the channel of collisional transport becomes more important.

The stabilization of core turbulence by nitrogen extends the region where neoclassical transport holds, and although the neoclassical peaking is reduced by the light impurity, it is still larger than the one driven by turbulence. Considering both collisional and turbulent transport channels, the tungsten peaking is enhanced in presence of the light impurity, and it matches bolometry measurements. 

The gyrokinetic prediction of the core tungsten peaking in WEST is validated with a synthetic diagnostic of the radiation power measured by the bolometers along the lines of integration (channels). Including nitrogen impurities is found to be essential for an accurate prediction, as their presence extends the region where neoclassical transport, which is subject to a strong peaking, dominates over the turbulent transport.

The possibility of getting an increased tungsten peaking in the core, as observed in this WEST experiment, has important implications for the plasma current ramp-up phase of ITER, where light impurities seeding{, such as Argon or Neon,} will be desirable to get low temperature at the plasma facing components and reduce tungsten sputtering. It provides further argument for applying early ECRH heating and get margins on the core power balance.

\section*{Acknowledgement}
The authors would like to thank T. Barbui, C. Bourdelle, L. Delgado, A. Diallo, F. Parra, and H. Zhu for discussions.

This research was supported by the SciDAC project ``Computational Evaluation and Design of Actuators for Core-Edge
Integration'' (CEDA) under contract number DE-FOA-0002924.

This research used INCITE resources of the Oak Ridge Leadership Computing Facilities at the Oak Ridge National Laboratory (OLCF) and Argonne National Laboratory (ALCF), which are supported by the Office of Science of the U.S. Department of Energy under Contract Nos. DE-AC05-00OR22725 and DE-AC02-05CH11231, respectively.

This research used resources of the National Energy Research Scientific Computing Center (NERSC), a Department of Energy Office of Science User Facility using NERSC award ERCAP0028234.

This work was supported by the U.S. Department of Energy under contract number DE-AC02-09CH11466. The United States Government retains a non-exclusive, paid-up, irrevocable, world-wide license to publish or reproduce the published form of this manuscript, or allow others to do so, for United States Government purposes. 

This work has been carried out within the framework of the EUROfusion Consortium, funded by the European Union via the Euratom Research and Training Program (Grant Agreement No 101052200 — EUROfusion). Views and opinions expressed are however those of the author(s) only and do not necessarily reflect those of the European Union or the European Commission. Neither the European Union nor the European Commission can be held responsible for them.

\section*{References}
\bibliographystyle{iopart-num}
\bibliography{bibliography}
\end{document}